\documentclass[12pt]{article}
\usepackage{graphicx}
\usepackage{cite}
\newcommand{\be}{\begin{equation}}
\newcommand{\ee}{\end{equation}}
\newcommand{\beq}{\begin{eqnarray}}
\newcommand{\eeq}{\end{eqnarray}}
\newcommand{\ba}{\begin{array}}
\newcommand{\ea}{\end{array}}
\newcommand{\crs}{cross section}
\newcommand{\crss}{\crs s}

\newcommand{\ppa}{\bar pp\to\eta\eta\eta}
\newcommand{\ppkk}{\bar pp\to\eta K\bar K}
\newcommand{\ppb}{\bar pp\to\eta K^0\bar K^0}
\newcommand{\ppc}{\bar pp\to\eta K^+ K^-}

\newcommand{\bfppa}{\mbox {\boldmath $\bar pp\to\eta\eta\eta$}}
\newcommand{\bfppc}{\mbox {\boldmath $\bar pp\to\eta K^+K^-$}}
\newcommand{\bfppb}{\mbox {\boldmath $\bar pp\to\eta K^0\bar K^0$}}
\newcommand{\bfekk}{\mbox {\boldmath $\bar pp\to\eta K\bar K$}}
\newcommand{\sss}{\!\!\!/}
\newcommand{\tql}{\textquotedblleft}

\newcommand{\dist}{\displaystyle}
\newcommand{\rpt}{\rule{0pt}{14pt}}
\newcommand{\rptt}{\rule{0pt}{16pt}}
\newcommand{\rttt}{\rule{0pt}{22pt}}

\newcommand{\mL}{\mathcal{L}}
\newcommand{\mD}{\mathcal{D}}
\newcommand{\half}{\frac{1}{2}}
\newcommand{\gamu}{\gamma_{\mu}}
\newcommand{\dga}{\gamma_{\mu}\gamma_5}
\newcommand{\uga}{\gamma^{\mu}\gamma_5}
\newcommand{\bfk}{{\bf k}}
\newcommand{\bfbb}{{\rm\bf B}}
\newcommand{\bfsig}{\mbox {\boldmath $\sigma$}}
\newcommand{\bfpi}{\mbox {\boldmath $\pi$}}
\newcommand{\bfll}{\mbox {\boldmath $L_{1-4}$}}
\newcommand{\bfbbb}{\mbox {\boldmath $B$}}
\newcommand{\bfma}{\mbox {\boldmath $M_A$}}
\newcommand{\bfmaa}{\mbox {\boldmath $M_{A_{1,2,3}}$}}
\newcommand{\bfmd}{\mbox {\boldmath $M_{D_{1,2}}$}}
\newcommand{\bfmf}{\mbox {\boldmath $M_{F_{1,2}}$}}

\newcommand{\dmu}{\partial_{\mu}}
\newcommand{\umu}{\partial^{\mu}}

\date{}

\begin{document}

\title{$\bar pp$-annihilation processes
       in the tree approximation of SU(3) chiral effective theory}

\author{V.E.~Tarasov$^1$, A.E.~Kudryavtsev$^1$, A.I.~Romanov$^{1,2}$,
        V.M.~Weinberg$^{1,3}$}
\maketitle

\vspace{-9mm}
\centerline{$^1$\it {Institute of Theoretical and Experimental Physics,
Moscow, 117218, Russia}}
\centerline{$^2$\it {National Research Nuclear University MEPhI,
Moscow,115409, Russia}}
\centerline{$^3$\it {Moscow Institute of Physics and Technology,
Moscow district, 141700, Russia}}


\vspace{-4mm}
\begin{abstract}
The ${\bar p}p$-annihilation reactions $\ppa$ and $\ppkk$ at rest are
considered in the tree approximation in the framework of $SU(3)$ chiral
effective theory at leading order. The calculated branchings are
compared with the data. The results for neutral ($\eta\eta\eta$,
$K^0\bar K^0\eta$) and charged ($K^+ K^-\eta$) channels are essentially
different.

$\\$
Pacs numbers: 12.39.Fe, 13.60.Le, 13.75.Cs, 25.43.+t.
\end{abstract}


\section{Introduction}

Nowadays, the QCD formulated in terms of quark and gluon fields
is confirmed as the fundamental theory of strong interactions.
However, the quantitative predictions for hadronic reactions meet with
unsolved problems coming from a non-perturbative character of the QCD
at large distances. The alternative effective field approach,
successfully applied in the low-energy hadron physics, is the so-called
Chiral Perturbation Theory (ChPT). This theory is based on the internal
$SU(3)_L\times SU(3)_R\times U(1)_V$ symmetry of QCD, violated by quark
masses, and is formulated in terms of hadrons -- mesons (as the Goldstone
bosons) and baryons. The ChPT, more than 30-year history of which can be
traced back to the pioneering works by Weinberg~\cite{We79} and Gasser
and Leutwyler~\cite{Ga84}, has developed into a powerful tool for
investigating the $\pi\pi$~\cite{Co}, $\pi N$~\cite{Ga88,Fe01} as
well as the few nucleon systems~\cite{Be01,Ep05}. A large number of
references can be found in the reviews~\cite{BKM,AM,Scherer,RM}.
The Lagrangian of the effective theory may generally contain infinite
number of terms with different number of derivatives and unknown
constants, and ChPT makes sense only at small particle
momenta ($p$) and meson masses ($\mu$), i.e., at
$p,\mu\ll\Lambda_{\chi}=4\pi f_{\pi}\sim 1\,$GeV, where
$f\approx 93$~MeV is the pion decay constant. In this case we have the
expansion parameter $\chi\sim\mu/\Lambda_{\chi}\ll 1$, and the
leading-order (LO) term of the Lagrangian is well defined.

In this paper we make the first attempt to apply the ChPT to the
annihilation processes $N\bar N\!\to$~mesons. We are interested in the
case of slow final mesons (with momenta $k\ll\Lambda_{\chi}$) in the
reaction rest frame. This situation takes place, for example, in the
multi-pion annihilation process $N\bar N\!\to\!n\pi$ with large
$n\sim\sqrt{s}/\mu\gg 1$.
The appropriate data up to $n=9$, obtained from $p\bar p$ annihilation
at rest can be found in~\cite{Kl}. Theoretical study of such
processes involves calculation of a large number of diagrams and
integration over the final many-particle phase space. In this respect,
the reactions with a small number of heavier final mesons $\eta$ and
$K(\bar K)$ seem to be more attractive, and corresponding data are
available as well~\cite{Ams}. In this paper, we consider the $p\bar p$
annihilation processes
\be
(a)~\bar pp\to\eta\eta\eta,~~~~
(b)~\bar pp\to\eta K^0\bar K^0,~~~~
(c)~\bar pp\to\eta K^+ K^-
\label{1}\ee
at rest. Here, the typical meson momentum is $k\sim 300\,$MeV/$c$, the
ChPT expansion parameter value $\chi\sim\mu/\Lambda_{\chi}\sim 0.5$,
and we may expect a sensible results from LO calculations in the
effective chiral theory. In the present paper, we perform calculations
for the probabilities (branchings) of the reactions~(\ref{1}) at rest
in the tree approximation, using the SU(3) chiral effective Lagrangian
at LO~\cite{AM,Scherer}.

The paper is organized as follows. In Section~2, we describe the
$SU(3)$ chiral Lagrangian and write down its parts used in the tree
calculations. In Section~3, we calculate the amplitudes for the
reactions~(\ref{1}). In Section~4, we present our numerical results
for the branchings, compare them with the experiment and discuss the
results. Section~5 is the Conclusion.

\section{Lagrangian}

We consider the $SU(3)$ chiral effective Lagrangian at leading order.
The meson part $\mL_{\pi\pi}$ of the Lagrangian can be written
in the form
\be
\mL_{\pi\pi}=\mL_1+\mL_2,
\label{LM}
\ee
$$
\mL_1=\frac{f^2}{4}\left\langle\dmu U\,\umu U^+\right\rangle,~~~
\mL_2=\sigma\frac{f^2}{2}\left\langle UM^++MU^+\right\rangle,~~~
U=\exp\left(\frac{2i\pi}{f}\right),
$$

$$
\pi=\half\sum\pi_a(x)\lambda_a =\frac{1}{\sqrt{2}} \left(\!
\ba{ccc}\dist
\frac{1}{\sqrt{2}}\,\pi^0\!+\frac{1}{\sqrt{6}}\,\eta &\dist\pi^+ &
\dist K^+\\ \dist
\pi^- &\dist -\frac{1}{\sqrt{2}}\,\pi^0\!+\frac{1}{\sqrt{6}}\,\eta &
\dist K^0\\
\dist K^- &\dist \bar{K}^0 &\dist -\frac{2}{\sqrt{6}}\,\eta
\ea\!\right).
$$
Here, $\pi$ is the octet matrix of the meson fields, and $U$ represents
their customary exponential parametrization; $\lambda_a$ ($a=1,...8$)
are the Gell-Mann matrices, normalized as
$\langle\lambda_a\lambda_b\rangle=2\delta_{ab}$; the angle brackets
$\langle\cdots\rangle$ denote the trace over flavours;
$f\approx 93$~MeV is the pion decay constant; $M=diag(m_u,m_d,m_s)$
is the quark-mass matrix.
The first term $\mL_1$ in Eq.~(\ref{LM}) is the chiral symmetrical part
of the Lagrangian $\mL_{\pi\pi}$. The second term $\mL_2$ contains the
quark-mass matrix $M$ and spontaneously violates the chiral symmetry.
The parameter $\sigma$ in the term $\mL_2$ relates the meson and quark
masses in the form
\be
\ba{ll}
m^2_{\pi}=\sigma (m_u\!+m_d),~~ &
m^2_{K^{\pm}}=\sigma (m_u\!+m_s),\\ \rptt
m^2_{K^0,\bar K^0}=\sigma (m_d\!+m_s),~~ & 
m^2_{\eta}=\frac{1}{3}\sigma (m_u\!+m_d\!+4m_s),
\ea
\label{mass}
\ee
known as the Gell-Mann, Oakes and Renner relations~\cite{GMOR}.$\!\!$
\footnote[4]{$^)$These also include the equation for the quark condensate
$\langle q\bar q\rangle=-3\sigma f^2$.}$^)$
To obtain Eqs.~(\ref{mass}) one can expand the Lagrangian
$\mL_2$~(\ref{LM}) in the meson fields and identify the quadratic term
for a given meson $\pi_a$ with the mass term $-\half m^2\pi^2_a$.

The baryon-meson part of the effective Lagrangian can be written as
\be
\mL_{\pi B}=\mL_3+\mL_4,
\label{LB}
\ee
$$
\mL_3=\langle\bar B\,(i\mD_{\!\mu}\gamma^{\mu}-m^{}_B)B\rangle,~~~
\mL_4=\langle\bar B\uga(D\{A_{\mu},B\}+F[A_{\mu},B])\rangle.
$$
where
\be
\ba{ll}
\mD_{\mu}=\dmu B+[V_{\mu},B],~~~ &
\dist V_{\mu}=\half(\xi\dmu\xi^+ +\xi^+\dmu\xi),
\\ \rttt\dist
A_{\mu}=\frac{i}{2}(\xi\dmu\xi^+ -\xi^+\dmu\xi),~~~ &
\dist \xi=\exp\left(\frac{i\pi}{f}\right)=\sqrt{U},
\ea
\label{LMB1}
\ee

$$
B=\frac{1}{\sqrt{2}}\sum B_a(x)\lambda_a =\left(\!
\begin{array}{ccc}\dist
\frac{1}{\sqrt{2}}\,\Sigma^0\!+\frac{1}{\sqrt{6}}\,\Lambda &
\Sigma^+ & p\\ \Sigma^- &\dist
-\frac{1}{\sqrt{2}}\,\Sigma^0\!+\frac{1}{\sqrt{6}}\,\Lambda & n\\
\Xi^- &\Xi^0 &\dist -\frac{2}{\sqrt{6}}\,\Lambda
\end{array}
\!\right).
$$
Here, $B$ is the octet matrix of the baryon fields; $m^{}_B$ is the
baryon mass; $D$ and $F$ are the usual axial-vector meson-baryon
coupling constants; $[A,B]$ and $\{A,B\}$ mean commutator and
anticommutator for the operators $A$ and $B$.

We expand the Lagrangians $\mL_{\pi\pi}$~(\ref{LM}) and
$\mL_{\pi B}$~(\ref{LB}) in meson fields and retain only the
interaction terms of interest for the $\bar pp\to$(3 mesons) amplitude
in the tree approximation. So, we use the terms of the fourth order in
meson fields from $\mL_{\pi\pi}$ and the terms up to the third order
from $\mL_{\pi B}$.
Then, the total interaction Lagrangian takes the form
\be
L=L_1+L_2+L_3+L_4,
\label{L}
\ee
$$
\ba{ll}\dist
L_1=\frac{1}{3f^2}\langle[\umu\pi,\pi][\dmu\pi,\pi]\rangle,~~ &
\dist L_2=\frac{2\sigma}{3f^2}\langle M\pi^4\rangle,
\\ \rttt\dist
L_3=i\langle\bar B\gamma^{\mu}[V_{\mu},B]\rangle,~~ &
L_4=\mL_4~(\rm see~Eq.~(\ref{LB})),
\\ \rttt\dist
V_{\mu}=\frac{1}{2f^2}[\pi,\dmu\pi], & \dist A_{\mu}=
  \frac{1}{f}\dmu\pi -\frac{1}{6f^3}[[\dmu\pi,\pi],\pi],
\ea
$$
Finally, since $m_{u,d}\ll m_s$, we take the $u$ and $d$ quarks to be
massless. Thus, we have $M=diag(0,0,m_s)$ and $\sigma=3m^2_{\eta}/4m_s$
(see Eqs.~(\ref{mass})) for the term $L_2$ in Eqs.~(\ref{L}).

\section{The amplitudes}

The full sets of the tree diagrams for the reactions~(\ref{1}a) and
(\ref{1}b), (\ref{1}c) are shown in Figs.~\ref{fig:1} and \ref{fig:2},
respectively.
Here we write down the corresponding amplitudes. The vertices of these
diagrams with a given mesons and baryons are fixed by the Lagrangians
$L_{1-4}$, and are given in Section~1 of Appendix.
We shall use the notations:
\be
k\sss\equiv k^{\mu}\gamu,~~
G_{\!B}(p)\!=\frac{p^{}\sss+m_B}{p^2\!-m^2_B\!+\!i0},~~
u=\!\sqrt{2m}\left(\!\!\ba{c}\varphi\\ 0\ea\!\!\right),~~
\bar v=\!\sqrt{2m}\,(0,-\chi^+).
\label{not}\ee
Hereafter, $G_{\!B}(p)$ is the propagator of baryon with 4-momentum $p\,$;
$u$($\bar v$) is the bispinor of the initial proton(antiproton) at rest;
$\varphi$ and $\chi$ are the spinors
($\varphi^+\!\varphi\!=\!\chi^+\!\chi\!=\!1$); $m$ is the proton mass.

\vspace{2mm}
\centerline{\bf 3.1.~The reaction $\bfppa$}
\vspace{2mm}

The amplitudes for the reaction~(\ref{1}a) read
\be
\ba{l}
iM_A=V^3_{p\bar p\eta}\,\bar v k\sss_3\gamma_5 G_{\!p}(k_3\!-\!p_2)
k\sss_2\gamma_5 G_{\!p}(p_1\!-\!k_1) k\sss_1\gamma_5 u
+ (k_1k_2k_3~ {\rm permutations}), \\ \rptt
iM_B=V_{p\bar p\eta}V^{(2)}_{\eta^4}XG_{\eta},~~~
X\!=\bar v P^{\!}\sss\gamma_5 u=-4m^2(\chi^+\!\varphi).
\ea
\label{a1}\ee
Here, $k_i\!=\!(\omega_i,\bfk_i)$ are the 4-momenta of the final
$\eta$ mesons; $p_1(p_2)\!=\!(m,\vec 0)$ is the 4-momentum of the
initial proton(antiproton) at rest, and $P\!=\!p_1\!+p_2$ is the
total 4-momentum; $G_{\!p}(\cdots)$ is the proton propagator
($B\!=\!p$, see Eq.(\ref{not}));
$G_{\eta}\!=\!1/(4m^2\!-\!m^2_{\eta})$ is the $\eta$ propagator in the
diagram ($B$); $V_{p\bar p\eta}$ and $V^{(2)}_{\eta^4}$ are the vertex
constants, given in Section~1 of Appendix.
The amplitude $M_A$~(\ref{a1}) contains $3!\!=\!6$ terms, arising from
permutations of $\eta$'s momenta $k_i$.

Further, we express the amplitudes $M$~(\ref{a1}) in the form
\be
iM=\chi^+(A+i\bfsig\bfbb)\varphi,
\label{a2}\ee
making use of Eqs.~(\ref{not}). Then for the diagrams
in Fig.~\ref{fig:1} we obtain
\be
\ba{l}
(A)\!:~ A,\bfbb = ({\rm see~Section~2~of~Appendix});
\\ \rptt
(B)\!:~ A=-4m^2 V_{p\bar p\eta}V^{(2)}_{\eta^4}G_{\eta},
~~ \bfbb=0.
\ea
\label{a3}\ee

We also consider the simplified amplitudes, calculated with final
mesons at rest. In this "threshold" $s$-wave approximation we take
\be
\bfk_i=0,~~~
\omega_i=\mu_i+\frac{T}{3},~~~ T=2m-\mu_1-\mu_2-\mu_3,
\label{a4}\ee
where $\mu_i$ are the meson masses, and $T$ is the total kinetic energy
(the final mesons are produced at rest as if their masses are
$\bar\mu_i=\omega_i$). In this case we have two graphs
\be
\ba{ll}\dist
(A)\!:~ A^{(0)}\!=\frac{3!\,V^3_{p\bar p\eta}\,
2m\omega^3_{\eta}}{(2m-\omega_{\eta})^2};~~~ &
(B)\!:~ A^{(0)}\!=A({\rm Eq.~(\ref{a3})}).
\ea
\label{a5}\ee
and the terms of the type $i\bfsig\bfbb$ vanish (we use a superscript
\tql(0)" for the amplitudes in the \tql threshold" approximation).

\vspace{2mm}
\centerline{\bf 3.2.~The reactions~ $\bfekk$}
\vspace{2mm}

The diagrams for these reactions are shown in Fig.~\ref{fig:2}.
Let us denote here the 4-momenta of final $\eta$, $K$ and $\bar K$ by
$k_1$, $k_2$ and $k_3$, respectively. Making use of the Lagrangians,
given in Section~1 of Appendix, we obtain the $\ppkk$ amplitudes in
the form
\be
\ba{l}
iM_{A_1}=V_{p\bar p\eta}V^2_{p\bar AK}\,\,\bar v k\sss_3\gamma_5 G_{\!A}
(k_3\!-\!p_2)k\sss_2\gamma_5 G_{\!p}(p_1\!-\!k_1) k\sss_1\gamma_5 u,
\\ \rptt
iM_{A_2}=V_{p\bar p\eta}V^2_{p\bar AK}\,\,\bar v k\sss_1\gamma_5 G_{\!p}
(k_1\!-\!p_2)k\sss_3\gamma_5 G_{\!A}(p_1\!-\!k_2) k\sss_2\gamma_5 u,
\\ \rptt
iM_{A_3}=V_{\!A\bar A\eta}V^2_{p\bar AK}\,\bar v k\sss_3\gamma_5 G_{\!A}
(k_3\!-\!p_2)k\sss_1\gamma_5 G_{\!A}(p_1\!-\!k_2) k\sss_2\gamma_5 u;
\ea
\label{b1}\ee
\be
\ba{l}
iM_B(1,a)=2V_{1a}\,
(2m\omega_{\eta}\!+m^2_K\!-m^2_{K\bar K})XG_a,
\\ \rptt
iM_B(2,a)=V_{2 a}XG_a,~~~ (a=\eta,\pi^0),~~~
V_{ia}=V_{p\bar pa}V^{(i=1,2)}_{a\eta K\bar K};
\ea
\label{b2}\ee

\be
\ba{ll}
~iM_C=-V_{\!C}\,\bar v (2k\sss_1\!-k\sss_2\!-k\sss_3)\gamma_5 u,~~~~~~~ &
V_{\!C}=V_{p\bar p\eta K\bar K};~~~~~
\ea
\label{b3}\ee

\be
\ba{ll}
~~~iM_{D_1}=-V_{\!D}\,
\bar v k\sss_3\gamma_5 G_{\!A}(k_3\!-\!p_2)(k\sss_2\!-\!k\sss_1) u, &
\\ \rptt
~~~iM_{D_2}=-V_{\!D}\,
\bar v (k\sss_1\!-\!k\sss_3)G_{\!A}(p_1\!-\!k_2) k\sss_2\gamma_5 u\,,
 & ~~V_{\!D}=V_{p\bar AK}V_{p\bar A K\eta}
\\ \rptt
~~~iM_{F_1}=-V_{\!F}\,
\bar v k\sss_1\gamma_5 G_{\!A}(k_1\!-\!p_2)(k\sss_2\!-\!k\sss_3) u, &
\\ \rptt
~~~iM_{F_2}=-V_{\!F}\,
\bar v (k\sss_2\!-\!k\sss_3)G_{\!A}(p_1\!-\!k_1) k\sss_1\gamma_5 u\,,
 & ~~V_{\!F}=V_{p\bar p\eta}V_{p\bar p K\bar K}.
\ea
\label{b4}\ee
Here, $m_{K\bar K}$ is the effective mass of $K\bar K$ system;
$G_a\!=(4m^2\!-\!m^2_a)^{-1}$ is the meson propagator ($a=\eta,\pi^0$),
and $m_a$ is the meson mass. The amplitudes~$M\!_B(i,a)$~(\ref{b2}) are
specified by the Lagrangian $L_i$ ($i=1,2$), determining the 4-meson
vertex of the diagram ($B$) in Fig.~\ref{fig:2}, and by the exchanged
meson $a=\eta,\pi^0$.
Representing the amplitudes in the form~(\ref{a2}), we obtain
\be
\ba{ll}
(A_{1,2,3}, D_{1,2}, F_{1,2})\!:~
 A,\bfbb =({\rm see~Sections~2,3~of~Appendix});
\\ \rptt
(B_{1,a})\!:~ A=-8V_{1a}m^2\,
(2m\omega_{\eta}\!+m^2_K\!-m^2_{K\bar K})G_a;
\\ \rptt
(B_{2,a})\!:~ A=-4V_{2a}m^2G_a;~~~
\\ \rptt
(C)\!:~ A=4V^{}_{\!C}\, m^2
(2\omega_{\eta}\!-\!\omega_K\!-\!\omega_{\bar K});~~
\bfbb\!=\!0~ {\rm for}~(B),(C);
\ea
\label{b5}\ee
In the \tql threshold" approximation~(\ref{a4}) we have $\bfbb\!=\!0$
and
\be
\ba{l}\dist
(A_{1,2})\!:~ A^{(0)}\!=
\frac{V_{p\bar p\eta}V^2_{p\bar AK}\,2m\,\omega_{\eta}\omega^2\!_K}
{(2m\!-\!m_{\eta})(M\!_A\!+\!m\!-\!m\!_K)};~~~
(A_3)\!:~ A^{(0)}\!=
\frac{V_{A\bar A\eta}V^2_{p\bar AK}\,2m\,\omega_{\eta}\omega^2\!_K}
{(M\!_A\!+\!m\!-\!m\!_K)^2};
\\ \rttt
(B_{1,a})\!:~ A^{(0)}\!=-8V_{1a}m^2\,
(2m\omega_{\eta}\!-\!3\omega^2\!_K)G_a;
~~
(B_{2,a})\!:~ A^{(0)}\!=({\rm see~Eq.~(\ref{b5}))};
\\ \rttt
(C)\!:~ A^{(0)}\!=8V^{}_{\!C}\,m^2(\omega_{\eta}\!-\!\omega_K);
~~ 
(D_{1,2})\!: \dist
A^{(0)}\!=V_{\!D}\frac{2m\omega\!_K(\omega\!_K\!-\!\omega_{\eta})}
{M\!_A\!+\!m\!-\!\omega\!_K};~~
(F_{1,2})\!: A^{(0)}\!=0.
\ea
\label{b6}\ee

\vspace{5mm}
\centerline{\bf 3.3.~A remark on power counting}
\vspace{2mm}

Let us look at the tree approximation in connection with the ChPT
power-counting rules and give the dimensional estimation of the
amplitudes. We express the amplitudes through the following factors:
$V$ (the product of the vertex constants); $2m$ (normalization factor
for bispinors, i.e. $\bar uu=2m$); $\omega^n$, where $\omega$ is
typical value of the meson total energy (for the meson-baryon vertex
of the diagrams ($B$) we replace $\omega\!\to\!2m$) and $n$ is the
number of derivatives in the Lagrangian, determining a given vertex);
$1/\omega$ and $1/4m^2$ for baryon and meson propagators, respectively.
Let us rewrite the factors $V$ through the dimensionless constants
$\bar V$. Then, $V=\bar V m^2_{\eta}/f^3\sim \bar V\omega^2/f^3$ for
diagrams ($B_{2,a}$) in Fig.~\ref{fig:2} (and diagram ($B$) in
Fig.~\ref{fig:1}), and $V=\bar V/f^3$ for all the rest of diagrams.
The resulting estimations of the diagrams in Figs.~\ref{fig:1} and
\ref{fig:2} are
\be
(A)\sim (C)\sim (D)\sim (F)\sim \frac{\bar V}{f^3}\,2m\omega,
~~ (B)\sim \frac{\bar V}{f^3}\,\omega^2.
\label{estim}\ee
Since $\omega\sim 2m/3\sim m$, the amplitudes~(\ref{estim}) are
approximately of the same order of magnitude. However, the $\bar V$
values, fixed by the $SU(3)$ Lagrangian, are very different
(these numbers are presented in Section~4 for the case of $K^+K^-\eta$
channel). Thus, the contributions of the tree diagrams are very
different, especially in the $\ppc$ case, as discussed below.

\section{The results}

Applying the amplitudes of Section~3, defined by Eq.~(\ref{a2}),
we get the $p\bar p$-annihilation \crs\ $v\sigma$ at rest ($v\to 0$)
in the form
\be
v\sigma(p\bar p\!\to\!\cdots)=\frac{I}{4m^2}\int\!
\overline{|M|^2} d\tau_3,~~~~~
d\tau_3=\frac{k_{1}q\,dwdz}{32\pi^3\sqrt{s}}~~~ (\sqrt{s}=2m).
\label{r1}\ee
Here, $I$ is the identity factor, and $I=1/3!$ ($I=1$) for the
$\eta\eta\eta$ ($\eta K\bar K$) channel;
~$\overline{|M|^{2}}=\half (|A|^{\,2}\!+\!|\bfbb|^{\,2})$ is the
amplitude~(\ref{a2}),
squared and averaged over the initial $p$ and $\bar p$ spin states;
$d\tau_3$ is the phase-space element (in the Feynman normalization)
of the final 3-meson state, written for the unpolarized case;
$k_1=Q(\sqrt{s},\mu_1,w)$ and $q=Q(w,\mu_2,\mu_3)$, where $Q(m,x,y)$
is the relative 3-momentum in the pair of particles with masses $x$
and $y$, and $m$ is the effective mass of this pair; $w$ is the
effective mass of the final meson pair (2$+$3), and $z=\cos\theta$,
where $\theta$ is the polar angle of the relative motion in the
(2$+$3) pair. The results obtained from Eqs.~(\ref{r1}) with the
amplitudes, given by Eqs.~(\ref{a3}) or (\ref{b5}), will be referred
to as the \tql exact" results.

We also give simplified (\tql threshold") version of calculations,
making use of the amplitudes, given by Eqs.~(\ref{a5}) and (\ref{b6}).
Then, the \crs\ can be written as
\be
v\sigma^{(0)}(p\bar p\!\to\!\cdots)=\frac{I}{8m^2}\,|A^{(0)\!}|^{\,2}
\,\tau^{(0)}_3,~~~~~ \tau^{(0)}_3=\frac{(\mu_1\mu_2\mu_3)^{1/2}\,T^2}
{64\pi^2(\mu_1\!+\!\mu_2\!+\!\mu_3)^{3/2}},
\label{r2}\ee
where $A^{(0)}$ is the sum of terms, given by Eqs.~(\ref{a5}) or
(\ref{b6}) for $\eta\eta\eta$ or $\eta K\bar K$ channels, respectively;
$\tau^{(0)}_3$ is the nonrelativistic phase space~\cite{Shap} of final
3-meson state, and $T=2m-\mu_1\!-\!\mu_2\!-\!\mu_3$ is the released
energy.

Further we give the results in terms of branchings (Br), where
\be
{\rm Br}(p\bar p\!\to\!\cdots)=
\frac{v\sigma(p\bar p\!\to\!\cdots)}{v\sigma^{\rm ann}_{p\bar p}}\,,
~~~~ v\sigma^{\rm ann}_{p\bar p}(v\!\to\!0)=36.5\,{\rm mb}.
\label{r3}\ee
The total $p\bar p$-annihilation \crs\ $v\sigma^{\rm ann}_{p\bar p}$
can be obtained from the imaginary part of $s$-wave scattering length
Im$^{}a_s=-0.69\,$fm, extracted in~\cite{CPZ}. Then, we get
$v\sigma^{\rm ann}_{p\bar p}(v\!\to\!0)=(8\pi/m)\,{\rm Im}(-a_s)=$
the value in Eqs.~(\ref{r3}).

Our numerical values of branchings as well as the experimental ones are
presented in Table~1. We give both \tql exact" and \tql threshold"
results (the latter are in brackets) for two sets -- $(a)$~\cite{Ok} and
$(b)$~\cite{Kub} -- of values of $D$ and $F$ constants, which enter the
term $L_4$ of the Lagrangian (\ref{L}) and were determined from
semileptonic baryon decays. These sets are
\be
\ba{l}
(a)~D=0.76,~ F=0.48~[17]; 
 \\ \rpt
(b)~D=0.80,~ F=0.46~[18]. 
\ea
\label{r4}\ee
The results in Table~1 show some sensitivity to $D$ and $F$ values.
Let us consider different channels separately.

\vspace{2mm}
\begin{center}
\begin{tabular}{l}
{\bf Table~1.}~ The branchings for $p\bar p$-annihilation reactions
at rest (the data~\cite{Ams}
\\
were obtained in liquid H$_2$ target; the data set for 3$\eta$ channel
was extracted
\\
from $p\bar p\to6\gamma$ and $p\bar p\to10\gamma$ events;
the $K^0\bar K^0\eta$ branching is obtained from
\\ $K_S K_S\eta$ data~\cite{Ams} as
Br$(K^0\bar K^0\eta)=2\,$Br$(K_S K_S\eta)$)
\end{tabular}

\vspace{2mm}
\begin{tabular}{|c|c|c|}
\hline\rpt Channel & Br(theor)$\times 10^4$ &
Br(exp)$\times 10^4$~\cite{Ams}
\\
\hline\rpt $\bar pp\to\eta\eta\eta$ &
$\ba{l}\rpt (a)~0.424~ (0.343)\\
            (b)~0.314~ (0.275) 
\ea$  &
$\ba{l}\rpt 4.22\pm 0.24~ (~6\gamma) \\ 3.93\pm 0.68~ (10\gamma)\ea$
\\
\hline\rpt $\bar pp\to\eta K^0\bar K^0$ &
$\ba{l}\rpt (a)~1.488~ (0.763)\\
            (b)~1.676~ (0.510) 
\ea$  &
$2\times (2.2\pm 0.7)$
\\
\hline\rpt $\bar pp\to\eta K^+ K^-$ &
$\ba{l}\rpt (a)~182.~ (0.0035)\\
            (b)~157.~ (0.0011) 
\ea$  & ------ \\
\hline\end{tabular}\end{center}

\vspace{1mm}
$\bfppa$. The calculated branchings underestimate the data~\cite{Ams}
approximately by one order of magnitude.
Let us consider our result ($a$) in more detail. The \tql exact" version
of branching value exceeds the \tql threshold" one by $\sim$ 24\%. The
diagrams ($A$) and ($B$) of Fig.~\ref{fig:1}, taken separately, give the
\tql exact" (\tql threshold") Br$(\eta\eta\eta)$ values $4.65\times 10^{-8}$
($1.16\times 10^{-6}$) and $4.50\times 10^{-5}$ ($4.81\times 10^{-5}$),
respectively. Thus, the diagram ($B$) dominates and the interference of
the diagrams ($A$) and ($B$) is destructive. The double-baryon-exchange
diagram ($A$) gives very different and relatively small contributions
in the \tql exact"/\tql threshold" versions. The amplitude $M_B$ is
approximately the same in both versions.$\!\!$
\footnote[5]{$^)$The difference of Br$(\eta\eta\eta)$ values from diagram
($B$) comes from the phase space difference $\tau_3/\tau^{(0)}_3=4.50/4.81$,
where $\tau^{(0)}_3$ is the non-relativistic phase space
in Eqs.~(\ref{r2}).}$^)$

\vspace{1mm}
$\bfppb$. Our results $(a)$, $(b)$ underestimate data by a factor
$\sim$~3. The experimental branching should also include the channels
with direct resonance production. One of them is the $\phi\eta$ channel,
studied in the reaction $\ppc$, and the extracted branching
Br$(^{3\!}S_1,\,p\bar p\to\phi\eta\to K^+K^-\eta)=
(0.76\pm 0.31)\!\times\! 10^{-4}$~\cite{Albe,Nomo,NS}, which should be
the same in the $K^0\bar K^0\eta$ channel. Subtracting this value with
weight factor $\frac{3}{4}$, we obtain
$(2\!\times\! 2.2-\frac{3}{4}\!\times\! 0.76=3.83)\!\times\! 10^{-4}$,
what is closer to our results. Some contribution should also come from
the $f_0(1500)\eta$ channel~\cite{Anis} due to the decay
$f_0(1500)\!\to\!K\bar K$~\cite{PDG}. With these reservations we may
accept the approach based on the diagrams of Fig.~\ref{fig:2}
as the \tql background" model for the $\ppb$ reaction at rest.

Contributions of the diagrams in Fig.~\ref{fig:2} to the branching
as well as signs of the cross terms are very different. The largest
contributions come from the diagrams ($B_{1,\,\pi^0}$), $(D_1\!+\!D_2)$
and $(F_1\!+\!F_2)$, and corresponding Br$(K^0\bar K^0\eta)\times 10^4$
values for variant $(a)$ are 2.30 (0.39), 3.09 (0.016) and 2.45 (0),
respectively. The results of \tql exact" and \tql threshold" versions
are drastically different, and the $(F_{1,2})$ terms vanish
in the latter case.

\vspace{1mm}
$\bfppc$. For this reaction the calculated branching exceeds the result
for Br$(K^0\bar K^0\eta)$ approximately by two orders of magnitude.
The experimental value for the charged channel case is unknown.
We roughly estimated it from the data on the annihilation
frequency $Y$~\cite{Nomo}. For the $s$-wave branching we may write
$Y>(1-f_p)\cdot Br(K^+K^-\eta)$, where
Br$(K^+K^-\eta)=\frac{1}{4}$Br$(^{1\!}S_0)+\frac{3}{4}$Br$(^{3\!}S_1)$,
and $f_p$ is the fraction of annihilation from the $p\,$-wave states.$\!\!$
\footnote[6]{$^)$Here, we neglect the correction factors $E(^{1\!}S_0)$
and $E(^{3\!}S_1)$ for the statistical weights $\frac{1}{4}$ and
$\frac{3}{4}$ of $^{1\!}S_0$ and $^{3\!}S_1$ states, equating them to 1.
For example, $E(^{3\!}S_1)=0.989$ for the liquid
$H_2$ target~\cite{Albe}.}$^)$
Taking the values $Y=(8.17\pm 0.37)\times 10^{-4}$~\cite{Nomo} and
$f_p=0.13$~\cite{Albe} for $\ppc$ at rest in the liquid target,
we obtain an estimation Br$(K^+K^-\eta)< 9.4\times 10^{-4}$ for the
upper value. Thus, the theoretical value exceeds the estimated data
by more than one order of magnitude.

It is worth noting that the \tql exact" version of our results exceeds
the \tql threshold" one approximately by five orders of magnitude, i.e.,
the latter approximation does not work at all. Very small Br$(K^+K^-\eta)$
values in the \tql threshold" version are due to cancellation of different
terms in the total amplitude. Table~2 contains the branching values,
obtained from different diagrams of Fig.~\ref{fig:2} separately. Here,
we give the \tql exact" and \tql threshold" results (the latter are in
brackets) for variant ($a$) and absolute values $|\bar V|$ of coupling
factors $\bar V$ (see Eq.~(\ref{estim})) for each diagram.
The subscripts $\Sigma^0$ and $\Lambda$ specifies the exchanged baryon~A
in Fig.~\ref{fig:2}. Further, we discuss the results of the \tql exact"
version.

The largest contributions come from the diagrams $(A_{1+2,\,\Lambda})$,
$(A_{3,\,\Lambda})$, $(D_{1+2,\,\Lambda})$ and $(F_{1+2})$. Note that
the terms $(D_{1+2,\,\Lambda})$ have the largest factor $|\bar V|=.238\,$
due to the large coupling constants, containing $\Lambda$ baryon.
With these terms excluded, i.e. for the incomplete set
$(A^{}_{1+2+3,\Sigma^0})+(B)+(C)+(D^{}_{1+2,\Sigma^0})$ of diagrams, we
have Br$(K^+K^-\eta)\times 10^4=6.2$. Successive addition of the terms
$(F_{1+2})$, $(A_{1+2+3,\,\Lambda})$ and $(D_{1+2,\,\Lambda})$ give
the values Br$(K^+K^-\eta)\times 10^4=12.1$, 42 and 182, respectively.
Note that the cross terms from these contributions are constructive.
Thus, we obtain a large branching value
Br$(K^+K^-\eta)\gg\,$Br$(K^0\bar K^0\eta)$, and the main reason comes
from the diagrams containing $\Lambda$ exchange, which are absent in the
neutral $K^0\bar K^0\eta$ channel.

\vspace{2mm}
\begin{center}
\begin{tabular}{l}
{\bf Table~2.}~ The branchings for $\ppc$ at rest from the diagrams \\
of Fig.~\ref{fig:2} (the subscripts $\Sigma^0$ and $\Lambda$ specifies
the exchanged baryon A)
\end{tabular}

\vspace{2mm}
\begin{tabular}{|c|c|c||c|c|c|}
\hline\rpt Diagram & Br$\times 10^4$ & $|\bar V|$ &
           Diagram & Br$\times 10^4$ & $|\bar V|$
\\ \hline
\rpt $(A_{1+2,\,\Sigma^0})$ & 0.0092 (0.0029) & 0.0038 &
     $(A_{3,\,\Sigma^0})$   & 0.0095 (0.0023) & 0.0086
\\ 
\rpt $(A_{1+2,\,\Lambda})$ & 4.57 (1.35) & 0.079  &
     $(A_{3,\,\Lambda})$   & 5.52 (1.19) & 0.177
\\
\rpt $(B_{1,\,\eta})$  & 0.91 (0.16) & 0.049 &
     $(B_{1,\,\pi^0})$ & 2.55 (0.46) & 0.089
\\
\rpt $(B_{2,\,\eta})$  & 0.44 (0.46) & 0.037 &
     $(B_{2,\,\pi^0})$ & 0.14 (0.14) & 0.045
\\
\rpt $(C)$  & 0.44 (0.15)   & 0.069 &
     $(D_{1+2,\,\Sigma^0})$ & 0.83 (0.0047) & 0.030
\\
\rpt $(D_{1+2,\,\Lambda})$ & 61.9 (0.32) & 0.238 &
     $(F_{1+2})$ & 10.6 (0) & 0.098
\\ \hline
\end{tabular}\end{center}

\section{Conclusion}

The $p\bar p$-annihilation processes $\ppa$ and $\ppkk$ at rest were
considered in the tree approximation in the framework of the $SU(3)$
chiral effective Lagrangian at leading order. The following results
were obtained.

\vspace{1mm}
1)The calculated branchings for the $\eta\eta\eta$ and $K^0\bar K^0\eta$
channels are several times smaller than the experimental values. Our
results underestimate the data approximately by one order of magnitude
in the $\eta\eta\eta$ case and -- by $\sim$ 3 times in the
$K^0\bar K^0\eta$ one.
One of the important missing contributions may be the direct production
of intermediate heavy resonances, not incorporated into the pseudoscalar
$SU(3)$ octet of mesons.
Thus, for example, an essential role of the resonance $f_0(1500)$ in the
$\eta\eta$-mass spectrum was found in the reaction $\ppa$ at beam
momenta 600-2410 MeV$/c$~\cite{Anis}. Due to a large $f_0(1500)$ width
$\Gamma\approx 109$~MeV~\cite{PDG}, one may also expect this resonance
to affect visibly the calculated \crs\ of the reaction at rest.
The other intermediate heavy resonances can also contribute.
On the other hand, we can not exclude a sizeable role of the
next-to-leading-order (NLO) corrections, including the loop diagrams.
Potential $p\bar p$-interaction in the initial state can also change the
reaction \crss\, calculated in the ChPT approximation [see, for example,
\cite{Shap1}].

Thus, we consider our estimation of the tree diagrams as the calculation
in the Born approximation and as the background model of the given
processes.

\vspace{1mm}
2)The calculated branching for the $K^+K^-\eta$ channel exceeds the
estimated data by more than one order of magnitude. Formally, the data
on this branching are unknown, and our statement is based on rough
estimation of the "experimental" value, given in Section~4.
Perhaps, the possible resonance-production mechanisms as well as the
NLO corrections could partly explain this strong discrepancy. However,
here the results obtained seem to be against the applicability of the
chiral effective approach to the annihilation processes. An important
reason of large calculated $K^+K^-\eta$ branching comes from the large
$SU(3)$ vertex constants, involving $\Lambda$-baryon exchange (not
occuring in the neutral $K^0\bar K^0\eta$ case). Perhaps, the NLO
corrections may be also enhanced due the $\Lambda$ vertices and improove
the theoretical branching value.
This is an opened question, faced to the further investigations.

\vspace{2mm}
\centerline{\bf Acknowlegements}
\vspace{1mm}

The authors are thankful to C.~Hanhart and V.G.~Ksenzov for useful
discussions. A.E.K. thanks grant NSh-4172.2010.2 for partial financial
support.

\vspace{3mm}
\centerline{\bf Appendix}
\vspace{2mm}



\centerline{\bf 1.~The terms of the Lagrangians $\bfll$~(\ref{L})
 for a given}
\centerline{\bf baryons ($\bfbbb$) and mesons ($\bfpi$)}

\vspace{2mm}
1.1.~The four-meson terms of $L_1$~(\ref{L}):
$$
\ba{l}
L_1(\eta\eta\,K\bar K)=V^{(1)}_{\eta\eta K\bar K}[\eta\dmu\eta\,
(K\umu\!\bar K\!+\bar K\umu\!K)-\dmu\eta\umu\!\eta\,K\bar K
\!-\eta\eta\,\dmu\!K\umu K],
\\ \rptt
L_1(\pi^0\eta K\bar K)=V^{(1)}_{\pi^0\eta K\bar K}[(\pi^0\dmu\eta\!
+\eta\,\dmu\pi^0)(K\umu\!\bar K\!+\bar K\umu\!K)-
\\ \rptt ~~~~~~~~~~~~~~~~~~~~~~~~~~~~~~~~~~~~~~~~~~~~~~~~~~~
 -2\dmu\eta\,\umu\!\pi^0 K\bar K\!-2\eta\pi^0\dmu K\umu\!\bar K],
\\ \rttt \dist
V^{(1)}_{\eta\eta K\bar K}=\frac{1}{4f^2},~~~
V^{(1)}_{\pi^0\eta K^+ K^-}= -V^{(1)}_{\pi^0\eta K^0\bar K^0}
=\frac{1}{4\sqrt{3}f^2}.
\ea
\eqno{(\rm A.1)}
$$

\vspace{1mm}
1.2.~The four-meson terms of $L_2$~(\ref{L}):
$$
\ba{l}
\dist
L_2(\pi_a\pi_b\pi_c\pi_d)=\frac{1}{n!}\,V^{(2)}_{\pi_a\pi_b\pi_c\pi_d}
\pi_a\pi_b\pi_c\pi_d,~~
\\ \rttt \dist
V^{(2)}_{\eta^4}=4!\,\frac{m^2_{\eta}}{18f^2},~~~
V^{(2)}_{\eta\eta K\bar K}=2!\,\frac{3m^2_{\eta}}{16f^2},~~~
V^{(2)}_{\pi^0\eta K^+ K^-}= -V^{(2)}_{\pi^0\eta K^0\bar K^0}
=-\frac{m^2_{\eta}}{8\sqrt{3}f^2}.
\ea
\eqno{(\rm A.2)}
$$

Here, we introduce the identity factor $n!$, where
$n=4(\eta^4),~ 2(\eta\eta K\bar K),~ 1(\pi^0\eta K\bar K).$

\vspace{3mm}
1.3.~The terms of $L_3$~(\ref{L}) for $B_a\bar B_b\to\pi_c\pi_d$:
$$\dist
L_3(B_a\bar B_b\pi_c\pi_d)=iV^{}_{B_a\bar B_b\pi_c\pi_d}\,
\bar u_b\gamu u_a\,(\pi_d\,\umu\pi_c-\pi_c\,\umu\pi_d),~~~~~~~~~
$$
$$
\ba{ll}\rttt\dist
V^{}_{p\bar\Sigma^+ K^0\eta}=-V^{}_{\Sigma^+\bar p\bar K^0\eta}=
\frac{\sqrt{3}}{4\sqrt{2}\,f^2}, &\dist
V_{p\,\bar\Lambda K^+\eta}=-V_{\Lambda\,\bar p K^-\eta}=
\frac{3}{8f^2},
\\ \rttt\dist
V_{p\,\bar\Sigma^0 K^+\eta}=-V_{\Sigma^0\bar p K^-\eta}=
\frac{\sqrt{3}}{8f^2}, &\dist
V_{p\,\bar p K^+K^-}\!=\frac{1}{2f^2},~~
V_{p\,\bar p K^0\bar K^0}=\frac{1}{4f^2};
\ea
\eqno{(\rm A.3)}
$$

\vspace{3mm}
1.4.~The terms of $L_4$~(\ref{L}) for $B_a\bar B_b\to\pi_c$:
$$
L_4(B_a\bar B_b\pi_c)=V^{}_{B_a\bar B_b\pi_c}\,\bar u_b\dga u_a\,
\umu\pi_a,~~~~~
V^{}_{B_a\bar B_b\pi_c}=V^{}_{B_b\bar B_a\bar\pi_c};~~~~~
$$
$$
\ba{lll}\dist
V^{}_{p\bar p\eta}=\frac{3F-D}{2\sqrt{3}\,f},~~ &\dist
V^{}_{p\bar p\pi^0}=\frac{D+F}{2f},~~ &\dist
V_{p\bar\Sigma^0 K^+}=\frac{D-F}{2f},~~
\\ \rttt\dist
V_{p\bar\Sigma^+ K^0}=\frac{D-F}{\sqrt{2f}},~~ &\dist
V_{p\bar\Lambda K^+}=-\frac{D+3F}{2\sqrt{3}f},~~ &\dist
V_{\Sigma\bar\Sigma\eta}=-V_{\Lambda\bar\Lambda\eta}
=\frac{D}{\sqrt{3}f};~~
\ea
\eqno{(\rm A.4)}
$$

\vspace{3mm}
1.5.~The terms of $L_4$~(\ref{L}) for $p\bar p\to\eta K\bar K$:
$$\dist
L_4(p\bar p\eta K\bar K)=V^{}_{p\bar p\eta K\bar K}\,
\bar u_{\bar p}\dga u_p\,
[2K\bar K\,\umu\eta-\eta (\bar K\umu K+K\umu\bar K)],
\eqno{(\rm A.5)}
$$

\vspace{-4mm}
$$\dist
V^{}_{p\bar p\eta K^0\bar K^0}=\frac{D-F}{8\sqrt{3}f^3},~~~~~~~
V^{}_{p\bar p\eta K^+\bar K^-}\!=\frac{-F}{4\sqrt{3}f^3};
$$

$u_p(\bar u_{\bar p})$ is the Dirac spinor of the proton (antiproton)
in items 1.3-1.5.

\vspace{5mm}

\centerline{\bf 2.~The amplitudes $\bfma$ and $\bfmaa$}
\vspace{2mm}

\begin{figure}
\begin{center}
\includegraphics[height=3.3cm, keepaspectratio]{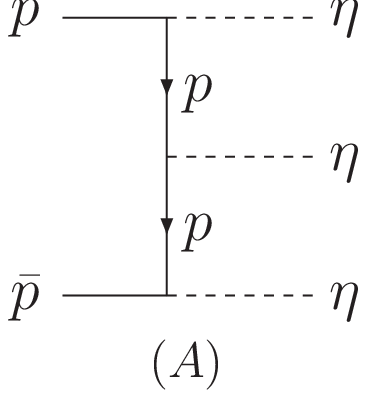}~~~~~~~~
\includegraphics[height=3.3cm, keepaspectratio]{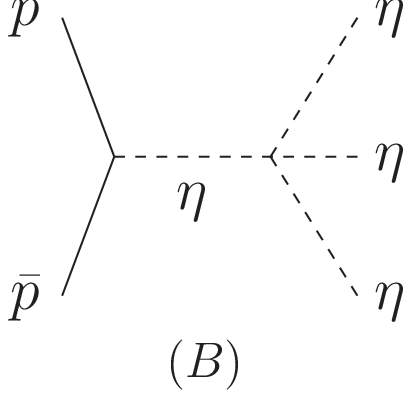}
\end{center}
\caption{Feynman tree diagrams for the $\bar p p\to\eta\eta\eta$
         amplitude. Solid and dashed lines correspond to the baryons
         and mesons, respectively.}\label{fig:1}
\end{figure}

\begin{figure}
\begin{center}
\includegraphics[height=3.45cm, keepaspectratio]{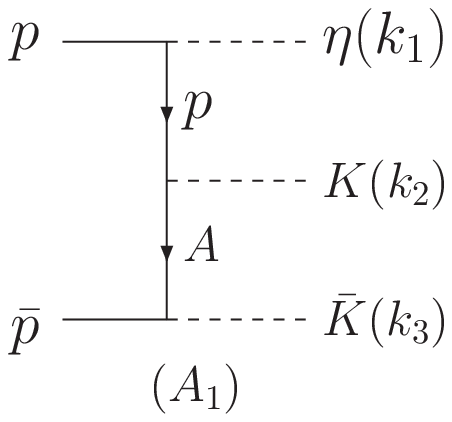}~~~~~~~~
\includegraphics[height=3.3cm, keepaspectratio]{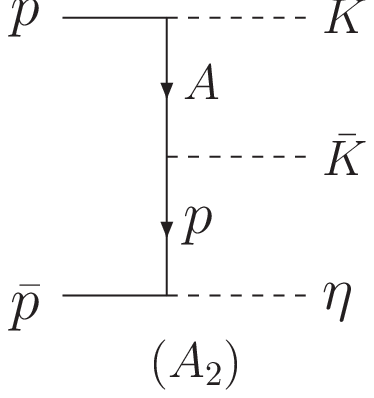}~~~~~~~~
\includegraphics[height=3.3cm, keepaspectratio]{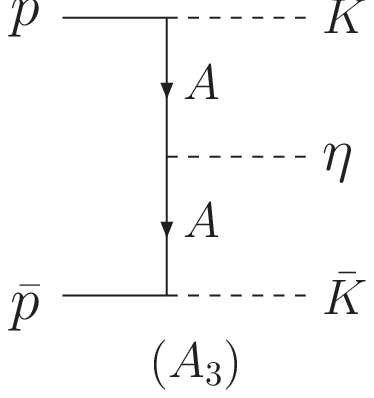}
\end{center}

\vspace{2mm}
\begin{center}
\includegraphics[height=3.3cm, keepaspectratio]{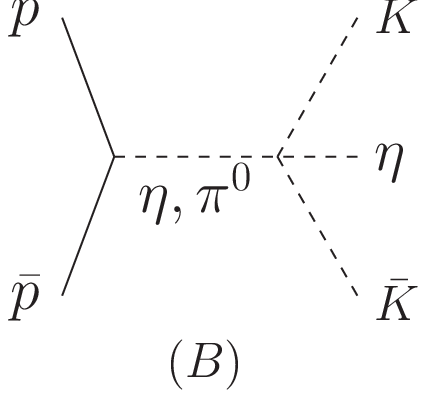}~~~~~~~~
\includegraphics[height=3.3cm, keepaspectratio]{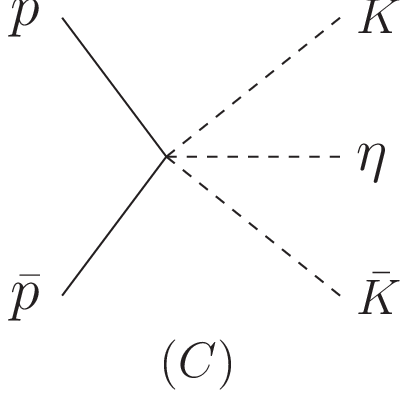}
\end{center}

\vspace{2mm}
\begin{center}
\includegraphics[height=3.3cm, keepaspectratio]{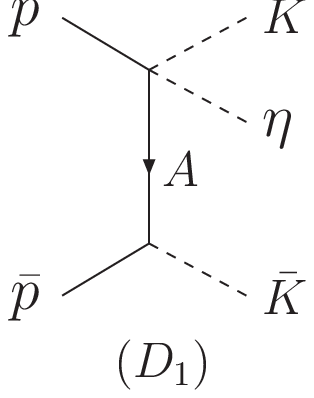}~~~~~~~
\includegraphics[height=3.3cm, keepaspectratio]{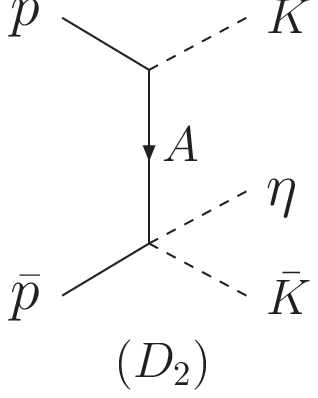}~~~~~~~
\includegraphics[height=3.3cm, keepaspectratio]{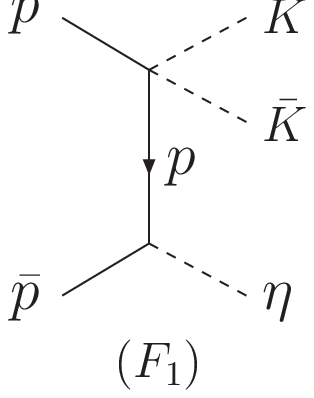}~~~~~~~
\includegraphics[height=3.3cm, keepaspectratio]{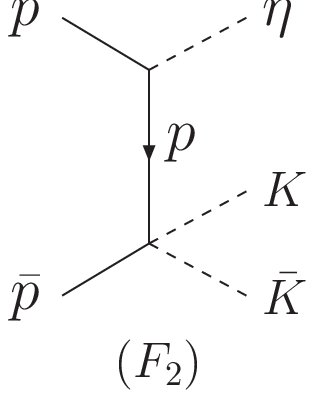}
\end{center}
\caption{Feynman tree diagrams for the $\bar p p\to\eta K\bar K$
amplitude. Lines are the same as in Fig~\ref{fig:1}. The exchanged
baryon $A$ is $\Sigma^+$ ($\Sigma^0$ or $\Lambda$) in the case of the
$\eta K^0\bar K^0$ ($\eta K^+ K^-$) channel.}\label{fig:2}
\end{figure}

For the amplitude $M_{A_1}$, represented in the form~(\ref{a2}),
we obtain
$$
(A_1)\!:~~ A=c[a+b(\bfk_1\bfk_3)],~~ \bfbb=cb\,[\bfk_3\times\bfk_1],
\eqno{(\rm A.6)}
$$
where
$$
c=2mV\,D^{-1}_1 D^{-1}_3,~~~
D_{1,3}=m^2\!-2m\omega_{1,3}\!+\mu^2_{1,3}\!-M^2_{1,3},
$$ $$
a=B_3(A_1 k^2_1\!+B_1\omega_2)+B_1 A_3 k^2_3,~~~
b=A_1(B_3\!+A_3\omega_2)+B_1 A_3,
$$ $$
A_{1,3}=m+M_{1,3},~~~
B_{1,3}\!=(M_{1,3}\!-m)\omega_{1,3}\!+\mu^2_{1,3}.
$$
Here: $V=V_{p\bar p\eta}V^2_{p\bar AK}$ is the vertex production;
$\bfk_i$, $\omega_i$ and $\mu_i$ are, respectively, the 3-momentum
($k_i=|\bfk_i|$), total energy and mass of the final meson
(see Fig.~\ref{fig:2}); $M_{1,3}$ are the exchanged-baryon masses
($M_1=m$ and $M_3=m_A$ for the diagram $A_1$ in Fig.~\ref{fig:2}).
The amplitudes $M_A$ (Fig.~\ref{fig:1}) and the rest of the amplitudes
$M_{A_i}$ (Fig.~\ref{fig:2}) can be obtained from Eq.~(A.6) by the
proper replacement of the vertex factor $V$, final-meson momenta and
baryon masses $M_i$.

\vspace{5mm}
\centerline{\bf 3.~The amplitudes $\bfmd$ and $\bfmf$}
\vspace{2mm}

For these amplitudes, represented in the form~(\ref{a2}), we obtain
$$
\ba{ll}
(D_{1,2})\!:~~ A=c_{1,2}[\,a_{1,2}\!+b(k^2_1\!-k^2_3)],~~ &
\bfbb=2c_{1,2\,} b\,[\bfk_3\times\bfk_2];
\\ \rptt
(F_{1,2})\!:~~ A=c[\,(\omega_{3,2}\!-\omega_{2,3})
\mu^2_1+2m(k^2_{3,2}\!-k^2_{2,1}],~~ &
\bfbb=4cm\,[\bfk_{1,3}\!\times\bfk_{2,1}].
\ea
\eqno{(\rm A.7)}
$$
Here,
$$
c_{1,2}=-2mV_D D^{-1}_{1,2},~~~~
D_{1,2}=m^2\!-2m\omega_{3,2}\!+\mu^2_{3,2}\!-m^{}_A,
$$ $$
a_{1,2}=(\omega_1\!-\omega_{2,3})
[(m^{}_A\!-m)\,\omega_{3,2}\!+\mu^2_{3,2}],~~~ b=m\!+m^{}_A,
$$ $$
c=-2mV_F(\mu^2_1\!-2m\omega_1)^{-1};$$
$V_D$ and $V_F$ are the vertex productions (\ref{b4}).

\newpage



\newpage

\centerline{\bf References}
\vspace{-15mm}
\begin{thebibliography}{99}
\bibitem{We79} S.~Weinberg, Physica A \textbf{96}, 327 (1979).
\bibitem{Ga84} J.~Gasser and H.~Leutwyler, Ann. Phys. \textbf{158}, 142 (1984).
\bibitem{Co} G.~Colangelo, J.~Gasser, and H.~Leutwyler, Nucl. Phys. B
             \textbf{603}, 125 (2001);\\ hep-ph/0103088.
\bibitem{Ga88} J.~Gasser, M.~E.~Sainio, and A.~Svarc, Nucl. Phys. B
               \textbf{307}, 779 (1988).
\bibitem{Fe01} N.~Fettes and U.-G.~Mei\ss ner, Nucl. Phys. A \textbf{693},
              693 (2001); hep-ph/0101030.
\bibitem{Be01} S.R.Beane, P.F. Bedaque, W.C.Haxton, {\it et al.},
  {\it At the Frontier of Particle Physics}
  (World Sci. Singapore, 2001), Vol. 1, p. 133; nucl-th/0008064.
\bibitem{Ep05}  E.~Epelbaum,
  Prog. Part. Nucl. Phys. \textbf{57}, 654 (2006); nucl-th/0509032.
%
\bibitem{BKM} V.~Bernard, N.~Kaiser, and Ulf-G.~Mei\ss ner,
         Int. J. Mod. Phys. E \textbf{4},193 (1995);\\ hep-ph/9501384.
\bibitem{AM} A.~Manohar, hep-ph/9606222. 
\bibitem{Scherer} S.~Scherer, Adv. Nucl. Phys. \textbf{27}, 277 (2003);
                  hep-ph/0210398.
\bibitem{RM} R.~Machleidt and D.~R.~Entem, Phys. Rept.
            \textbf{503}, 1 (2011); arXiv:1105.2919 [nucl-th].
\bibitem{Kl} E.~Klempt, Ch.~Batty, and J.-M.~Richard, Phys. Rept.
            \textbf{413}, 197 (2005);\\ hep-ex/0501020.
\bibitem{Ams} C.~Amsler {\it et al.}, Nucl. Phys. A
             \textbf{720}, 357 (2003).
\bibitem{GMOR} M.~Gell-Mann, R.~J.~Oakes, and B.~Renner, Phys. Rev.
              \textbf{175}, 2195 (1968).
\bibitem{Shap} I.~S.~Shapiro, UFN \textbf{92}, 549 (1967).
\bibitem{CPZ} J.~Carbonel, K.~V.~Protasov, and A.~Zenoni, Phys. Lett. B
             \textbf{397}, 345 (1997);\\ nucl-th/9806032.
\bibitem{Ok} L.~B.~Okun, {\it Leptons and Quarks} (Nauka, Moscow, 1990),
             Chapt. 6, p. 51.
\bibitem{Kub} B.~Kubis, hep-ph/0703274.
\bibitem{Albe} The OBELIX Collab. (A.~Alberico {\it et al.}),
               Phys. Lett. B \textbf{432}, 427 (1998).
\bibitem{Nomo} The OBELIX Collab. (V.~Nomokonov), Acta Phys. Polon. B
        \textbf{29}, 2547 (1998).
\bibitem{NS} V.~P.~Nomokonov and M.~G.~Sapozhnikov, Phys. Part. Nucl.
            \textbf{34}, 94 (2003) $\\$ [Fiz. Elem. Chastits At. Yadra
            \textbf{34}, 189 (2003)].
\bibitem{Anis} A.~V.~Anisovich {\it et al.}, arXiv:1109.4008 [hep-ex].
%
\bibitem{PDG} K.~Nakamura {\it et al.} (Particle Data Group), J. Phys. G
         \textbf{37}, 075021 (2010)\\ (available at http://pdg.lbl.gov).
\bibitem{Shap1} I.~S.~Shapiro, Phys. Rept. \textbf{35}, 129 (1978).
\end{thebibliography}
\end{document}